\documentclass[proceedings]{JHEP3} 

\PrHEP{PrHEP hep2001}                                   
\conference{International Europhysics Conference on HEP}

\usepackage{epsfig}
\def\epem{e$^+$e$^-$}
\def\Mepem{\mathrm{e^+e^-}}
\def\Mmpmm{\mu^+\mu^-}
\def\ccbar{$\mathrm{c}\overline{\mathrm{c}}$}
\def\dahad{\Delta\alpha_{\mathrm{had}}^{(5)}}
\def\Mwa{\omega_a}
\def\afbb{A_{\mathrm{FB}}^{\mathrm{b}}}
\def\PZz{Z$^0$}
\def\sintwl{\sin^2\theta_{\mathrm{eff}}^{\mathrm{lept}}}
\def\ipb{pb$^{-1}$}
\def\ppbar{p$\overline{\mathrm{p}}$}
\def\Oalpha{\mathcal{O}(\alpha)}
\def\mH{m_{\mathrm{H}}}
\def\Wqqqq{\mathrm{W}\mathrm{W}\to\mathrm{q}\mathrm{q}\mathrm{q}\mathrm{q}}
\def\Wqqlv{\mathrm{W}\mathrm{W}\to\mathrm{q}\mathrm{q}\ell\nu}
\def\Wlvlv{\mathrm{W}\mathrm{W}\to\ell\nu\ell\nu}
\def\etal{{\it et al.}}
\def\mtop{m_{\mathrm{top}}}
\def\mw{m_{\mathrm{W}}}
\def\mz{m_{\mathrm{Z}}}
\def\mh{m_{\mathrm{H}}}

\title{Experimental Tests of the Standard Model}

\author{\speaker{David Charlton}\\
        Royal Society University Research Fellow\\
        School of Physics \& Astronomy\\
        The University of Birmingham \\
        BIRMINGHAM B15 2TT, UK\\
        E-mail: \email{Dave.Charlton@cern.ch}}

\abstract{The current status of experimental tests of the
          electroweak sector of the Standard Model is 
          reviewed.}

\begin{document}

\vspace*{-11cm}
\begin{flushright}
BHAM-HEP/01-02\\
31 October 2001
\end{flushright}
\vspace*{9cm}

\section{Introduction}

The field of precise experimental tests of the electroweak sector of
the Standard Model encompasses a wide range of experiments.
The current status of these is reviewed in this report, with
emphasis placed on new developments in the year preceding
summer 2001.
A theme common to many measurements is that theoretical and
experimental uncertainties are comparable.
The theoretical uncertainties, usually coming from the lack of
higher-order calculations, can be at least as hard to estimate
reliably as the experimental errors.

At low energies, new hadronic cross-section results in \epem\
collisions are discussed.
The new measurement of the muon anomalous magnetic moment at
Brookhaven is reported and compared with recent Standard Model
calculations.
Results from the now complete LEP data sample are reviewed, together
with recent results from the Tevatron, HERA and SLD.
The synthesis of many of these results into a global test of the
Standard Model via a comprehensive fit is summarised.
Finally, prospects for the next few years are considered.

Many results presented here are preliminary: they are not labelled 
explicitly for lack of space. References should be
consulted for details.

\section{R and $\mathbf{\alpha(M_Z^2)}$}

\EPSFIGURE{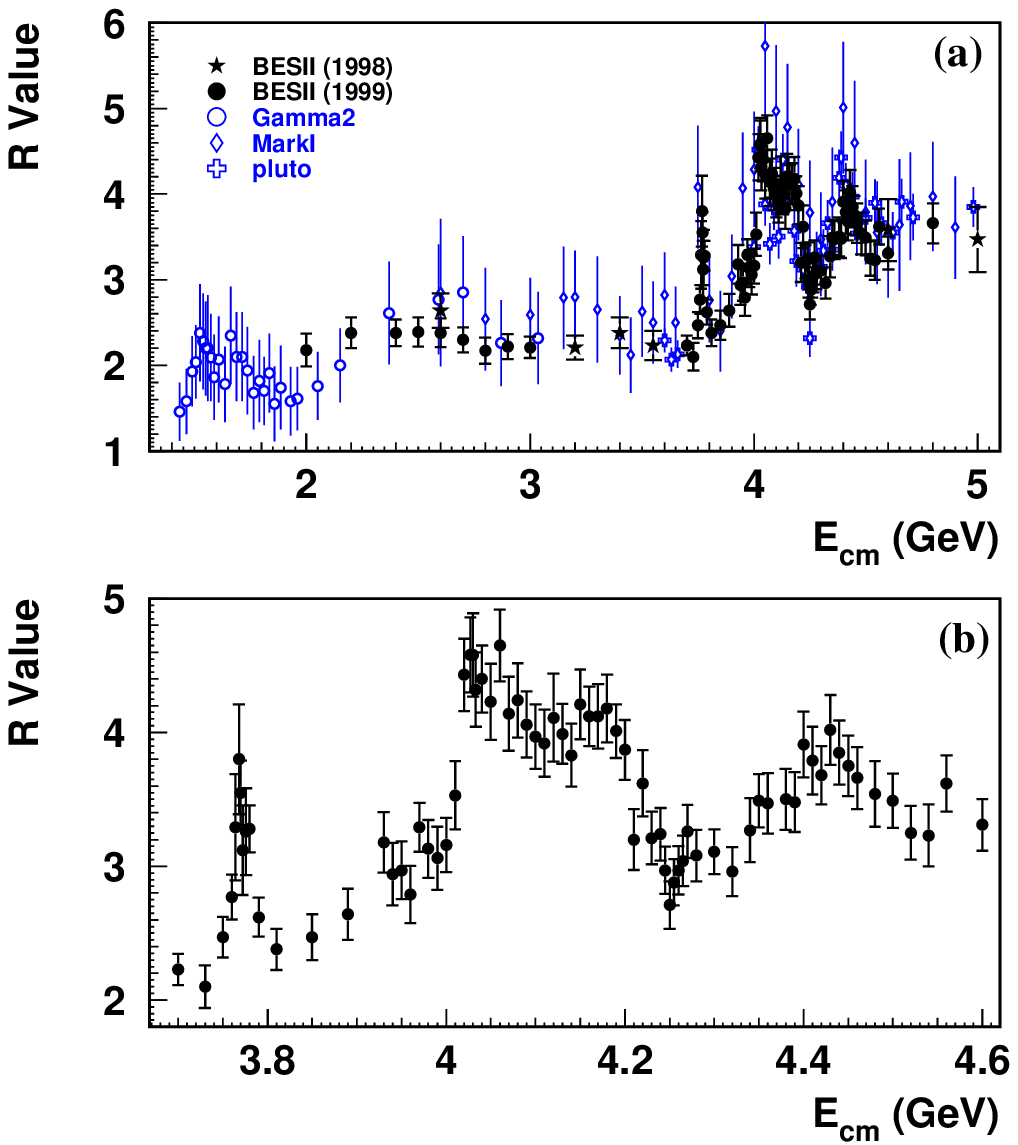}
{Measurements of R from BES: (a) Over the full $\sqrt{s}$ range; (b)
in the \ccbar\ resonance region\cite{bib:besr}.\label{fig:besr}}

The BES-II detector at the BEPC electron-positron collider in Beijing,
China, has been operating since 1997. 
Many measurements have been made in the centre-of-mass energy range
$2<\sqrt{s}<5$~GeV, but of relevance to electroweak physics are
those of the ratio
\[
R = \frac{\sigma(\Mepem\to\mathrm{hadrons})}{\sigma_0(\Mepem\to\Mmpmm)}
\]
where the denominator,
$\sigma_0(\Mepem\to\Mmpmm)=4\pi\alpha^2(0)/(3s)$, 
is the lowest-order QED prediction.
The BES measurements~\cite{bib:besr} of R are presented in 
Figure~\ref{fig:besr},
where the improvement in quality over previous, often very early,
measurements is clear. 
Around 1000 hadronic events are used at each
energy, and an average precision of 6.6\% is obtained at each
of the 85 energy points. The point-to-point correlated error is
estimated to be 3.3\%, providing a factor of 2 to 3 improvement over
earlier measurements.

In order to achieve such an improvement, detailed studies of the
detector
acceptance for hadronic events at low $\sqrt{s}$ were made, in
collaboration with the Lund Monte Carlo team. 
The experimental acceptance for hadronic events varies in the range
50 to 87\% from 2 to 4.8~GeV respectively, so the modelling at low
$\sqrt{s}$ is of most concern.
Good descriptions of the
hadronic event data were obtained from a tuned version of the {\tt
LUARLW} generator, and the hadronic model-dependent uncertainty is 
estimated to be as low as 2-3\%.

At even lower energies, analysis continues of the large data sample
from CMD-2~\cite{bib:cmd2} 
at the VEPP-2M \epem\ collider at Novosibirsk taken over
$0.36<\sqrt{s}<1.4$~GeV. Many exclusive final-states are studied, with
the main contribution to the overall cross-section arising from
$\pi^+\pi^-$ production.

A key application of the low energy R measurements is in the
prediction of the value of the electromagnetic coupling at the \PZz\ mass
scale. This is modified from its zero-momentum value,
$\alpha(0)=1/137.03599976(50)$, by vacuum polarisation loop corrections:
\[
\alpha(M_Z^2)=\frac{\alpha(0)}{1-\Delta\alpha_{e\mu\tau}(M_Z^2)-
\dahad(M_Z^2)-\Delta\alpha_{top}(M_Z^2)} .
\]
The contributions from leptonic and top quark loops
($\Delta\alpha_{e\mu\tau}$ and $\Delta\alpha_{top}$, respectively) 
are sufficiently well calculated knowing only the particle masses.
The $\dahad$ term contains low-energy hadronic loops, and
must be calculated via a dispersion integral:
\[
\dahad(M_Z^2) = - \frac{\alpha
M_Z^2}{3\pi} \Re\int_{4m_{\pi}^2}^\infty ds
\frac{R(s)}{s(s-M_Z^2-i\epsilon)} .
\]
The R data points must, at least, be interpolated to evaluate this
integral. 
More sophisticated methods are employed by different authors, and use
may also be made of $\tau$ decay spectral function data via
isospin symmetry.
A recent calculation~\cite{bib:pietrzyk} 
using minimal assumptions has
obtained
$\dahad(M_Z^2)=0.02761\pm0.00036$, approximately a factor two more
precise than a previous similar estimate which did not use the new
BES-II data.
With extra theory-driven assumptions, an error as low as $\pm0.00020$
may be obtained~\cite{bib:martin}.

Prospects for further improvements in measurements of the hadronic
cross-section at low energies are good: an upgraded accelerator in
Beijing should give substantially increased luminosity; CLEO proposes
to run at lower centre-of-mass energies than before to examine the
region from 3 to 5 GeV; DA$\Phi$NE may be able to access the low
energy range with radiative events; and finally the concept of a very
low energy ring to work together with the present PEP-II LER could 
give access to the poorly covered region between 1.4 and 2 GeV.

\section{The Muon Anomalous Magnetic Moment g-2}

\EPSFIGURE{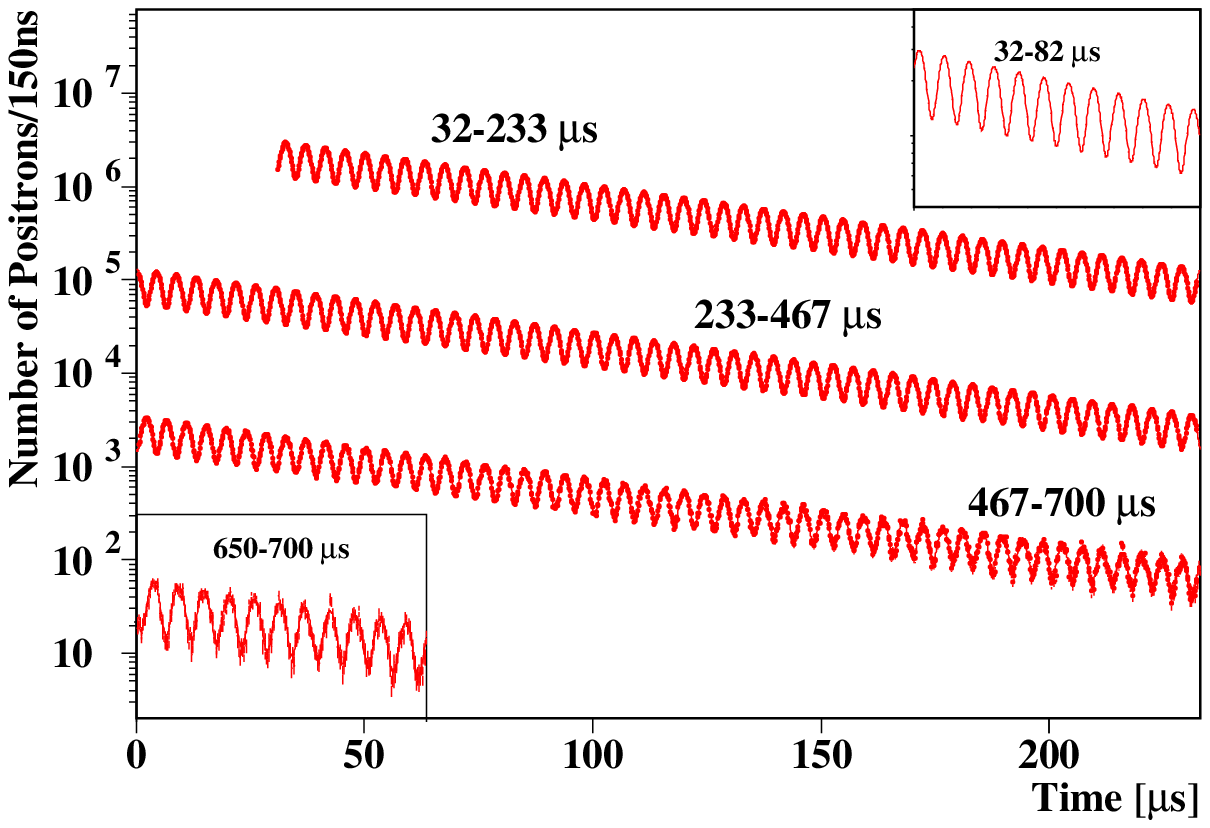}                              
{Time spectrum of positrons observed with energy $E>2$GeV in
E821\cite{bib:e821}. 
The periodicity from spin precession is observed over more than ten
muon lifetimes.\label{fig:e821}} 

The Brookhaven E821 experiment has recently reported~\cite{bib:e821} 
a new measurement of the muon anomalous magnetic moment, $a_{\mu}$,
by measuring the spin-precession
frequency, $\Mwa$, of polarised muons in a magnetic field:
\[
a_{\mu} \equiv \frac{g-2}{2} = \frac{\Mwa m_{\mu}c}{e\langle B\rangle}
\]
The muons circulate in a special-purpose storage ring constructed to
have an extremely uniform magnetic field across its aperture.
The spin-precession frequency $\Mwa$ is measured by observing the time
variation of production of decay
electrons above a fixed energy cut-off (2~GeV), 
as shown in Figure~\ref{fig:e821}.
The mean bending field is measured using two sets of NMR probes:
one fixed set mounted around the ring and used for
continuous monitoring, and another set placed on a trolley which can
be pulled right around the evacuated beam chamber.
In practice, the magnetic field is re-expressed in terms of the mean
proton NMR frequency, $\omega_p$, and $a_{\mu}$ extracted from:
\[
a_{\mu} = \frac{R}{\lambda-R}
\] 
where $R=\omega_a/\omega_p$ and $\lambda$ is the ratio of muon to
proton magnetic moments.

The latest E821 result, obtained using $0.95\times10^9$ $\mu^+$ decays is~\cite{bib:e821}:
\[
a_{\mu^+} = (11\,659\,202\pm14\pm6)\times 10^{-10}
\]
The overall precision obtained is relatively 1.3 parts per million:
1.2 ppm from statistics and 0.5 ppm from systematic errors. 
Data from a further $4\times10^9$
$\mu^+$ and $3\times10^9$ $\mu^-$ are in hand, and should result in a
factor two improvement in the near future.

Interpretation of this result in terms of the Standard Model and
possible new physics requires detailed calculations of loop
corrections to the simple QED $\mu\mu\gamma$ vertex, which 
gives the original $g=2$ at lowest order.
The corrections may be subdivided into electromagnetic (QED), weak and
hadronic parts according to the type of loops.
The QED and weak terms are respectively calculated to be $a_{\mu}(QED)
= (11\, 657\,470.57 \pm  0.29) \times 10^{-10}$, and $a_{\mu}(weak)
= (15.2 \pm 0.4) \times 10^{-10}$.
The hadronic corrections, although much smaller than the QED
correction,
provide the main source of uncertainty on the predicted
$a_{\mu}$. To $\mathcal{O}(\alpha^3)$, the dominant corrections 
may be subdivided into the lowest and higher-order vacuum 
polarisation terms and higher-order ``light-on-light'' terms.
The lowest-order (vacuum polarisation) term is numerically much the
largest. It can be calculated using a dispersion relation:
\[
a_{\mu}(had;LO) = \frac{\alpha^2(0)}{3\pi^2}\int_{4m_{\pi}^2}^{\infty}
ds \frac{R(s)\hat{K}(s)}{s^2}
\]
where $\hat{K}(s)$ is a known bounded function.
As for $\alpha(M_Z^2)$, optional additional theory-driven
assumptions may be made.
Recent estimates of the lowest-order vacuum polarisation term are
shown in Table \ref{tab:lovp}. 
There is some ambiguity at the level of $\sim$5$\times10^{-10}$ 
about the treatment of further photon radiation in some of these 
calculations, as it may be included either here or
as a higher-order correction, depending also on whether the input 
experimental data includes final-states with extra photons.
The estimates agree with each other within the overall errors, which
is not surprising since the data employed is mostly in common. 
It is notable that the best value available at the time of the E821
publication was that of Davier and H\"ocker (``DH(98/2)''), 
which is numerically the lowest of the calculations.

\TABLE{
\begin{tabular}{l|l|c}
Authors                           & Based on      & $a_\mu$(LO; hadronic) / 10$^{-10}$ \\
\hline
BW(96) \cite{bib:bw96}    & \epem\ data   &   703 $\pm$ 16 \\
ADH(98) \cite{bib:adh98}   & \epem\ data   &   695 $\pm$ 15 \\
ADH(98) \cite{bib:adh98}   & \epem\ \& $\tau$ data &   701 $\pm$ 9 \\
DH(98/1) \cite{bib:dh981} & \epem\ \& $\tau$ data &   695 $\pm$ 8 \\
DH(98/2) \cite{bib:dh982} & \epem\ \& $\tau$ data, QCD sum rules & 692 $\pm$
6 \\
N(01) \cite{bib:narison}   & \epem\ \& $\tau$ data & 702 $\pm$ 8 \\
J(01) \cite{bib:jegerlehner}&\epem\ \& $\tau$ data & 699 $\pm$ 11 \\
dTY(01) \cite{bib:dty01}   & \epem\ \& $\tau$ data & 695 $\pm$ 6 \\
CLS(01) \cite{bib:cls01} & QCD+renormalons \& data & 700 $\pm$ 9 \\
\end{tabular}
\caption{Recent calculations of the lowest-order hadronic correction
to $a_{\mu}$. 
\label{tab:lovp}}}

\EPSFIGURE{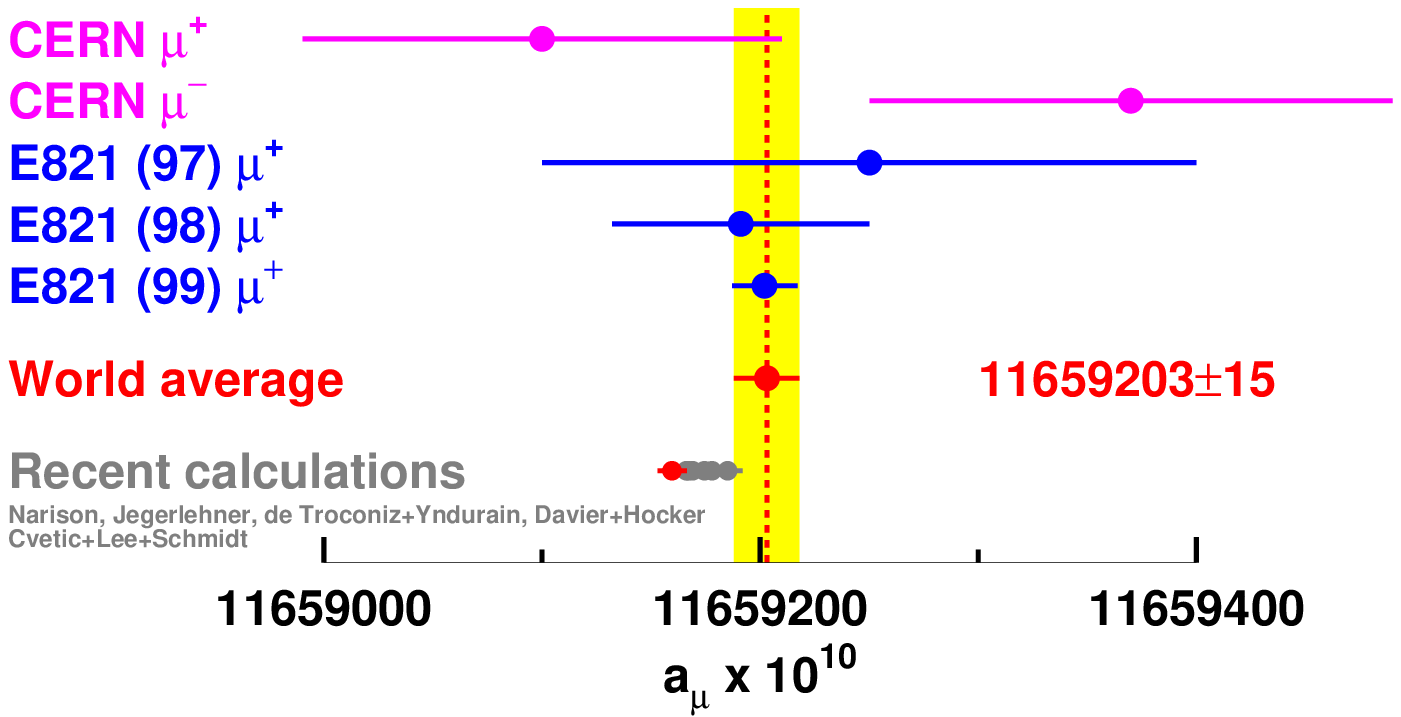}
{Measurements of the muon anomalous magnetic moment and recent
predictions.\label{fig:gminus2}} 

The summed corrections are shown in figure~\ref{fig:gminus2} and
compared with the new and previous measurements~\cite{bib:oldgm2}. 
The major experimental improvement from the new E821 measurement is
striking.
At the time of publication, the most precise available calculation of
$a_{\mu}$ led to a difference between data and theory of around 2.6
standard deviations~\cite{bib:e821}. 
More recent calculations reduce that difference, in some cases to
the one standard deviation level, thus also suggesting that the
error on the prediction may have been too optimistic.
At present there is therefore no reason to consider $a_{\mu}$ as
giving evidence of physics beyond the Standard Model.
The accuracy of the theoretical predictions will be even more
severely challenged by an experimental measurement with a factor two
smaller error, as expected in the near future. 
Theoretical progress is essential to obtain a maximum physics return
from such a precise measurement.

\section{Recent News from the Z$^{\mathbf{0}}$ Pole}
\label{sec:asy}

Measurements of the \PZz\ cross-section, width, and asymmetries have 
been available for many years from LEP and SLD data, and most results
have now been finalised~\cite{bib:lepls,bib:alr}.
Recently new results~\cite{bib:afbbnew} have become available on the b
quark
forward-backward asymmetry ($\afbb$) from ALEPH and DELPHI, using
inclusive lifetime-based b-tagging techniques and various quark charge 
indicators.
Substantial improvements are obtained over earlier lifetime-tag
measurements, so that this type of asymmetry measurement now has
a comparable precision to that using a traditional lepton tag.
The lepton and lifetime results are compatible, and together give 
a LEP average Z pole asymmetry of
\[
A_{\mathrm{FB}}^{\mathrm{0,b}} = 0.0990 \pm 0.0017 .
\]

\EPSFIGURE{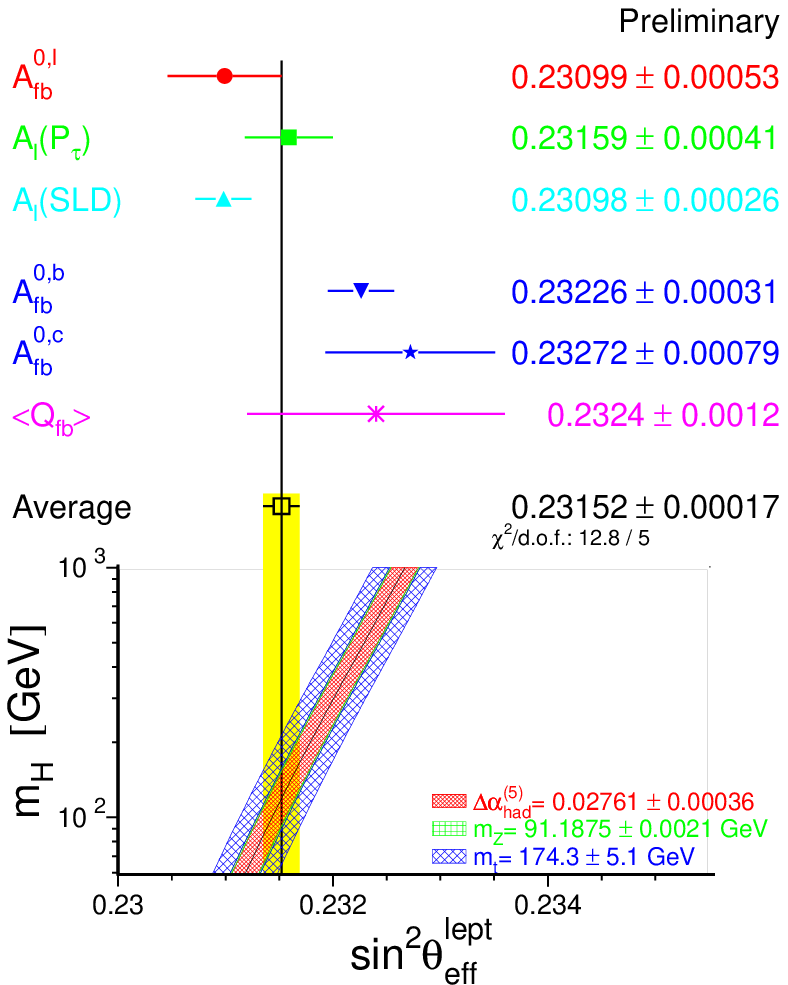}
{Comparison of asymmetry measurements interpreted simply as
measurements of $\sintwl$.
\label{fig:sstw}} 

This result may be compared with other asymmetry measurements from
LEP and SLD by interpreting $\afbb$ in terms of $\sintwl$.
In doing this, it is effectively assumed that the b quark couplings
are given by their Standard Model values.
The result is shown in figure~\ref{fig:sstw}, comparing to $\sintwl$
values derived from the leptonic forward-backward asymmetry from LEP
($A_{\mathrm{fb}}^{\mathrm{0,l}}$)~\cite{bib:lepls}; that from the
$\tau$ polarisation
measurements ($P_{\tau}$)~\cite{bib:ptau}; from the left-right
polarisation
asymmetry at SLD~\cite{bib:alr}; 
from the charm forward-backward asymmetry~\cite{bib:afbc}; 
and from inclusive hadronic
event forward-backward asymmetry measurements
($Q_{\mathrm{fb}}$)~\cite{bib:qfb}.
The two most precise determinations of $\sintwl$, from $A_{\mathrm{LR}}$ and
$\afbb$, differ at the level of 3.2 standard deviations. 
This might suggest that the b quark couplings to the \PZz\ differ from 
the Standard Model expectations, but such an interpretation is not 
compelling at present, and direct measurements via the left-right
polarised forward-backward b quark asymmetry at SLD are not precise
enough to help.
Future improvements in b quark asymmetry measurements using the
existing LEP data samples may help elucidate this issue, but scope for
such improvement is limited.

\section{LEP-2 and Fermion-Pair Production}

\TABLE{
\begin{tabular}{c|c}
Model         & Limit (TeV) \\
\hline
$\chi$        & 0.678 \\
$\psi$        & 0.463 \\
$\eta$        & 0.436 \\
LR            & 0.800 \\
Sequential    & 1.89 \\
\end{tabular}
\caption{95\% CL lower limits on the mass of new Z$^\prime$
bosons in various models~\cite{bib:ffmeas}.\label{tab:fflim}}}

With the completion of LEP-2 data-taking at the end of 2000, the
integrated luminosity collected at energies of 161~GeV and above has 
reached 700~\ipb\ per experiment, in total giving each 1~fb$^{-1}$ 
from the entire LEP programme.
Following on from the measurements of the LEP-1 Z lineshape and
forward-backward asymmetries,
studies of fermion-pair production have continued at LEP-2.
At these higher energies, fermion-pair events may be subdivided into
those where the pair invariant mass has ``radiatively returned'' to
the Z region or below, and non-radiative events with close to the 
full centre-of-mass energy.
The cross-sections and forward-backward asymmetries for non-radiative
events at the full range of LEP-2 energies are shown in
Figures~\ref{fig:fflep1} and~\ref{fig:fflep2} for hadronic, muon and
tau pair final states, averaged between all four LEP
experiments~\cite{bib:ffmeas}. 
Analogous measurements have been made for electrons, b and c
quarks~\cite{bib:ffmeas}.
The Standard Model expectations describe the data well.
Limits can be placed on new physics from these data~\cite{bib:ffmeas}. 
As an example, limits may be placed on new Z$^\prime$
bosons which do not mix with the \PZz, as indicated in
Table~\ref{tab:fflim}.

\DOUBLEFIGURE{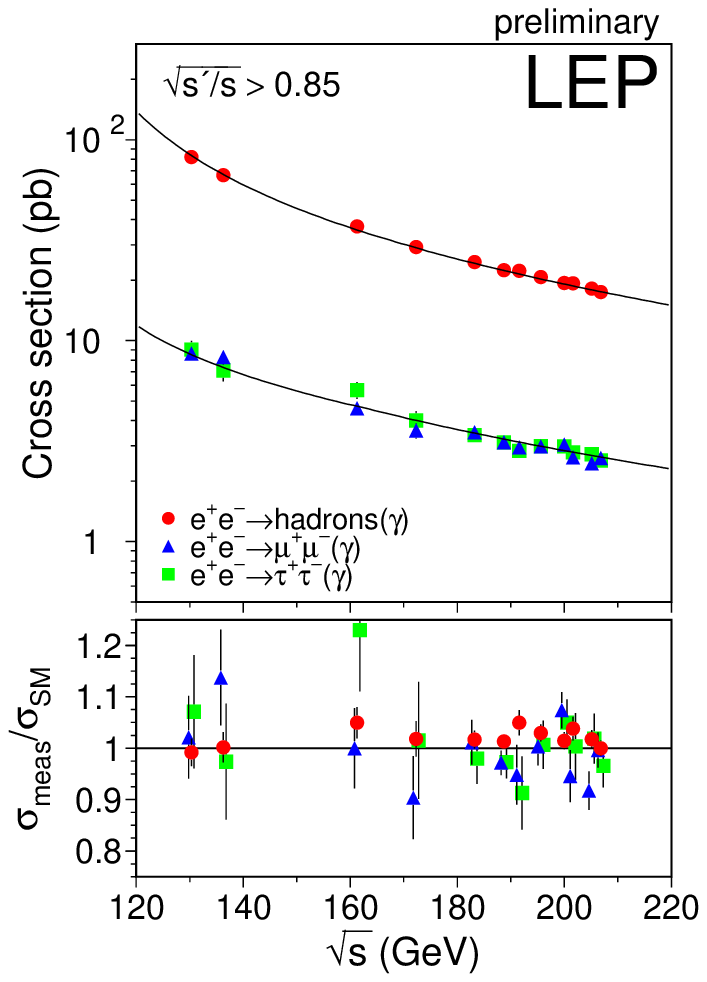}{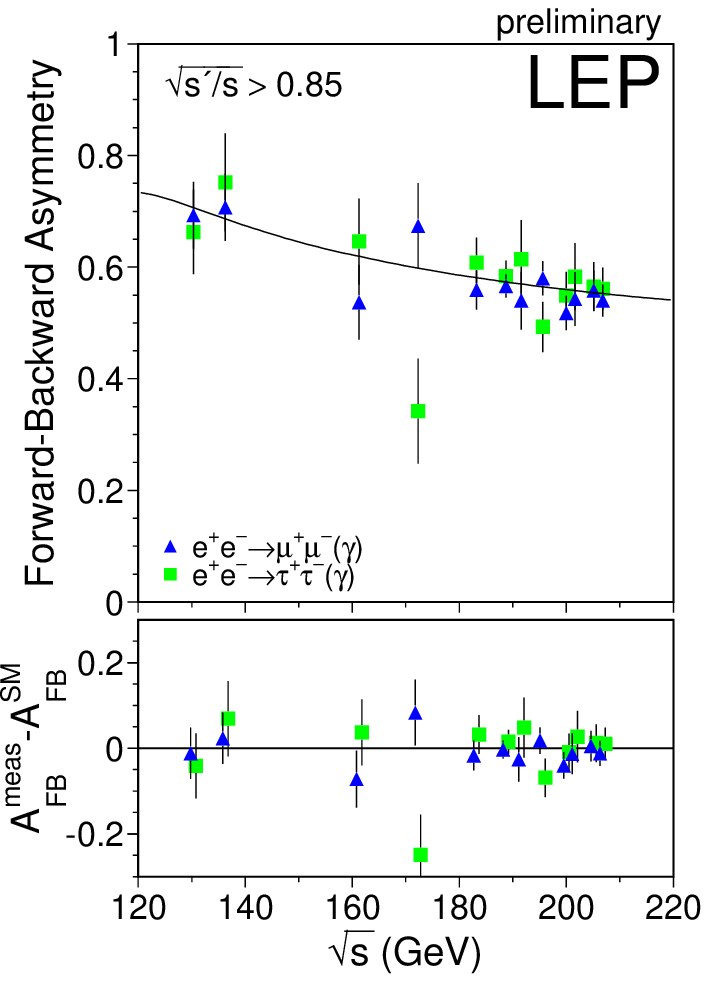}
{LEP combined fermion pair
cross-sections in the LEP-2 energy region\cite{bib:ffmeas}.\label{fig:fflep1}}
{LEP combined fermion pair
forward-backward asymmetries in the LEP-2 energy
region\cite{bib:ffmeas}.\label{fig:fflep2}}

\section{Z's and W's at Colliders with Hadrons}

\EPSFIGURE{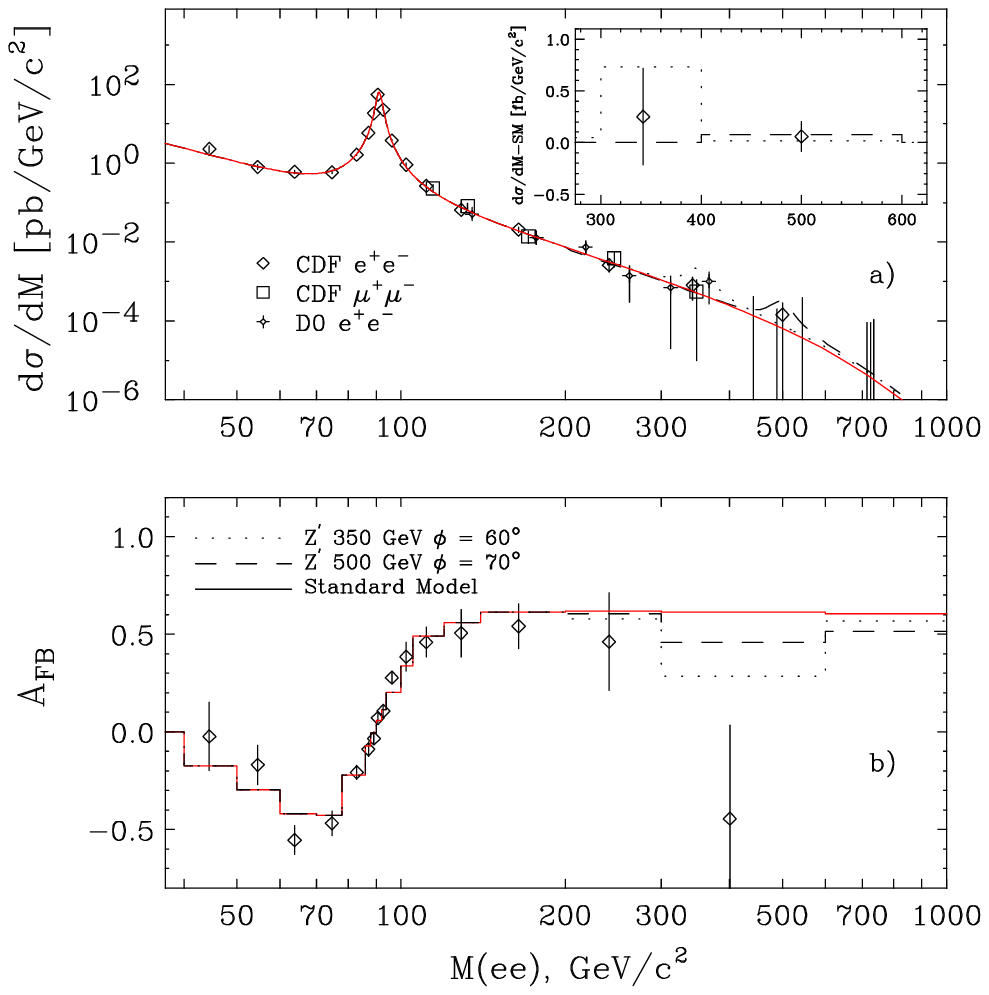}
{Measurements of Drell-Yan lepton-pair production by CDF.
\label{fig:drellyan}} 

Electroweak fermion-pair production has also been studied at the
Tevatron, in the Drell-Yan process.
Updated results on high mass electron pairs were presented at this
conference~\cite{bib:cdfdy,bib:gerber}: both cross-sections and
asymmetries 
are well described by
the Standard Model expectations, and extend beyond the LEP-2 mass
reach to around 500 GeV (see Figure~\ref{fig:drellyan}).
As indicated in the figure, there is some sensitivity to new physics
models, and improvements on that of LEP should come with the Run 2
data.

\TABLE{
\begin{tabular}{cc|c}
Experiment    & Sample & Mass (GeV) \\
\hline
UA2           & W~$\to$~e$\nu$ & 80.360$\pm$0.370 \\
CDF           & Run 1A         & 80.410$\pm$0.180 \\
CDF           & Run 1B         & 80.470$\pm$0.089 \\
D0            & Run 1A         & 80.350$\pm$0.270 \\
D0            & Run 1B         & 80.498$\pm$0.084 \\
\hline
\multicolumn{2}{c|}{All \ppbar\ data}& 80.454$\pm$0.060 \\
\end{tabular}
\caption{W mass measurements from hadron colliders.\label{tab:mwhad}}}

W production in \ppbar\ collisions provided, before LEP-2, the only
direct measurements of the W mass, using reconstructed electron and
muon momenta and inferred missing momentum information.
The main results from CDF and D0 from Run 1 data have been available
for some time~\cite{bib:mwtev}.
D0 have recently updated their Run 1 results with a new analysis
making use of electrons close to calorimeter cell
edges~\cite{bib:gerber}. 
The main importance of
the extra data is to allow a better calorimeter calibration from Z
events.
Measurements of the W mass from the Tevatron are summarised in
Table~\ref{tab:mwhad}.

The high tail of the distribution of the transverse mass of the
lepton-missing momentum system provides information about the W
width. 
CDF finalised their Run 1 result ($\Gamma_{\mathrm{W}}=$~2.05$\pm0.13$
GeV)~\cite{bib:gwcdf} some time ago.
D0 presented a new measurement using all the Run 1 data, of
$\Gamma_{\mathrm{W}}=$~2.23$\pm0.17$ GeV, at this
conference~\cite{bib:gerber}.

\DOUBLEFIGURE{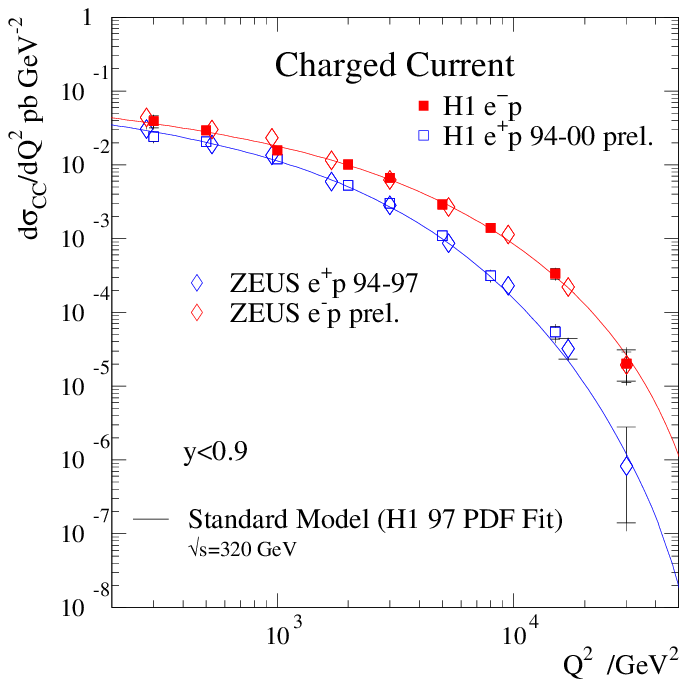}{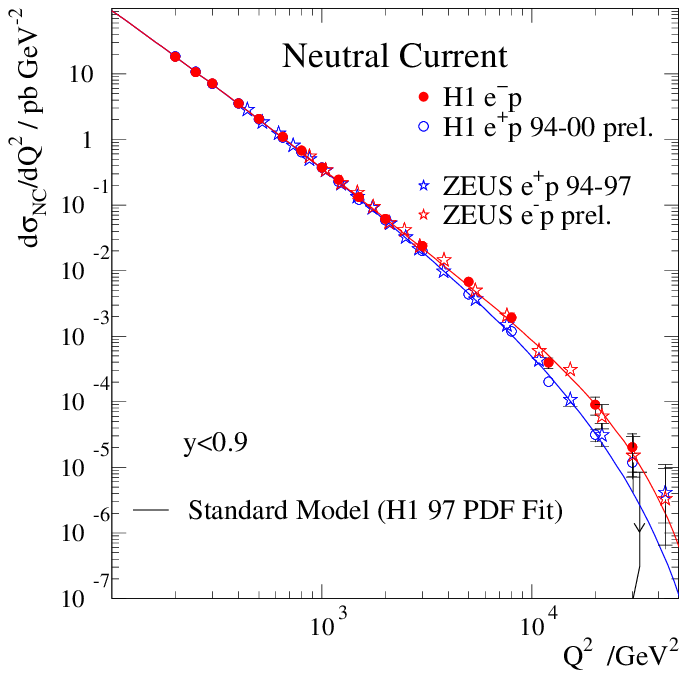}
{Charged-current differential cross-sections measured at
HERA\cite{bib:heracc}.\label{fig:heracc}}
{Neutral-current differential cross-sections measured at
HERA\cite{bib:heranc}.\label{fig:heranc}}

The presence of the W and Z bosons is primarily probed at HERA via
t-channel exchange.
The charged and neutral current differential cross-sections as a
function of $Q^2$ are shown in Figures~\ref{fig:heracc}
and~\ref{fig:heranc} respectively.
The charged current process proceeds only by W exchange, and is
sensitive to the W mass via the propagator term (and also, indirectly,
via the overall normalisation).
The effect of \PZz\ exchange can be seen in the high-$Q^2$ neutral
current region where it gives rise to a difference between the e$^-$p
and e$^+$p cross-sections.

Real W production may also have been observed at HERA, by looking for
events with high transverse momentum electrons or muons, missing
transverse momentum, and a recoiling hadronic system.
For transverse momenta of the recoiling hadronic system above 40 GeV,
H1 and ZEUS together observe 6 events compared to an
expectation of 2.0$\pm$0.3, which is 90\% composed of W production and
decay~\cite{bib:whera}.
These events have been interpreted as possible evidence of new
physics, but within the framework of the Standard Model their natural
interpretation is as W production.

\section{W Physics at LEP-2}

\EPSFIGURE{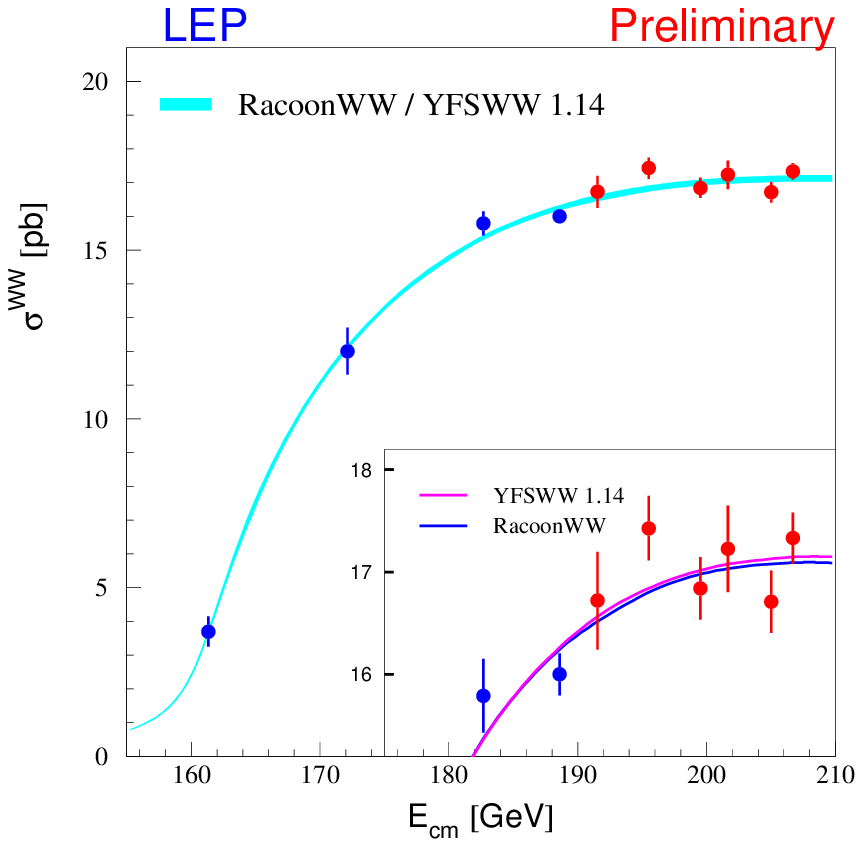}{
LEP averaged W-pair production cross-section measurements\cite{bib:sigww}.
\label{fig:sigww}} 

Each LEP experiment now has a sample of around 12000 W-pair
events from the full LEP-2 data sample.
Event selections are well established, and have needed only minor
optimisations for the highest energy data. 
Typical selection performances give efficiencies and purities in the
80-90\% range for almost all channels -- channels with $\tau$ decays
being the most challenging. 
The measured W-pair cross-section~\cite{bib:sigww} 
is shown in Figure~\ref{fig:sigww},
and compared to the predictions of the RacoonWW~\cite{bib:racoon} and
YFSWW~\cite{bib:yfsww} Monte Carlo programs. 
These programs incorporate full $\Oalpha$ corrections to the
doubly-resonant W-pair production diagrams, and give a
cross-section approximately 2\% lower than earlier
predictions. 
The agreement can be tested by comparing the experimental and
predicted cross-sections as a function of centre-of-mass energy.
The new calculations describe the normalisation of the data well,
the old ones over-estimate it by between two and three standard
deviations of the experimental error~\cite{bib:sigww}.

The selected W-pair events are also used to measure the W decay
branching ratios. 
The combined LEP results~\cite{bib:sigww} are shown in
Table~\ref{tab:wbr}. 
The leptonic results are consistent with lepton universality, and so
are combined to measure the average leptonic branching ratio,
corrected to massless charged leptons. 
This measurement now has a better than 1\% relative error, and is
consistent with the Standard Model expectation of 10.83\%.
It is significantly more precise than a value extracted from the
Tevatron W and Z cross-section data, assuming Standard Model 
production of W's, which is
Br(W$\to\ell\nu)=10.43\pm0.25$\%~\cite{bib:wbrtev}.

\DOUBLETABLE{\begin{tabular}{c|c}
Decay mode    & Branching ratio (\%)\\
\hline
W~$\to$~e$\nu$ & 10.54$\pm$0.17 \\
W~$\to\mu\nu$  & 10.54$\pm$0.16 \\
W~$\to\tau\nu$ & 11.09$\pm$0.22 \\
W~$\to\ell\nu$ & 10.69$\pm$0.09 \\
\hline
W~$\to$~hadrons& 67.92$\pm$0.27 \\
\end{tabular}}
{\begin{tabular}{l|c}
Sample               & W Mass (GeV)\\
\hline
ALEPH (1997-2000)    & 80.477$\pm$0.038$\pm$0.032\\
DELPHI (1997-2000)   & 80.399$\pm$0.045$\pm$0.049\\
L3 (1997-2000)       & 80.389$\pm$0.048$\pm$0.051\\
OPAL (1997-1999)     & 80.491$\pm$0.053$\pm$0.038\\
\hline
$\Wqqlv$             & 80.448$\pm$0.033$\pm$0.028\\
$\Wqqqq$             & 80.457$\pm$0.030$\pm$0.054\\
\hline
From $\sigma_{\mathrm{W}\mathrm{W}}$(161 GeV) & 80.40$\pm$0.21 \\
\hline
LEP combined         & 80.450$\pm$0.039 \\
\end{tabular}}{W decay branching ratio measurements from
LEP\cite{bib:sigww}.\label{tab:wbr}}
{W mass measurements from LEP~\cite{bib:mwlepex,bib:mwlep}. 
Results are from the direct
reconstruction technique unless indicated.\label{tab:lepmw}}

The W mass and width are measured above the W-pair threshold at LEP-2
by direct reconstruction of the W decay products~\cite{bib:mwlepex}, 
using measured
lepton momenta and jet momenta and energies. Events with two
hadronically decaying W's (``$\Wqqqq$''), or where one W decays 
hadronically and the
other leptonically (``$\Wqqlv$''), are used by all experiments. 
A kinematic fit is
made to the reconstructed event quantities, constraining the total
energy and momentum to be that of the colliding beam particles, thus
reconstructing the unobserved neutrino in mixed hadronic-leptonic
decay events.
This fit significantly improves the resolution on the W mass.
The reconstructed mass distributions can be fitted to obtain the W
mass, or the W mass and width together. 
Other, more complicated,
techniques to extract the most W mass information from the fitted
events are used by some experiments.
ALEPH and OPAL also use the small amount of information contained in
$\Wlvlv$ events, which has been included in the $\Wqqlv$ results
quoted.

After the kinematic fit, the W mass statistical sensitivity is
very similar for the two event types. The systematic error sources are
largely different between the two channels: the main correlated
systematics come from the knowledge of the LEP beam energy, and
hadronisation modelling.
The W mass measurements obtained by the four LEP experiments, and
averaged by channel, are shown in table~\ref{tab:lepmw}. There is good
consistency between all the measurements, and the overall
precision~\cite{bib:mwlep} 
now improves significantly on the 60~MeV from hadron colliders.
If the W width is also fitted, the W mass measurement is essentially
unchanged, and a LEP combined value of
$\Gamma_{\mathrm{W}}=2.150\pm0.091$~GeV is found.

The 39~MeV error on the combined LEP result includes 26~MeV
statistical and 30~MeV systematic contributions. 
Systematic errors are
larger in the $\Wqqqq$ channel (see Table~\ref{tab:lepmw}), having the
effect of deweighting that channel, to just 27\%, in the
average. 
With no systematic errors this deweighting would not
occur, and the statistical error would be 22~MeV.
The main systematic errors on the combined result are as
follows~\cite{bib:mwlep}: The LEP beam energy measurement contributes
a highly
correlated 17~MeV to all channels; hadronisation modelling
uncertainties contribute another 17~MeV; ``final-state interactions''
(FSI) 
between the hadronic decay products of two W's contribute 13~MeV; 
detector-related uncertainties -- different for the different
experiments -- contribute 10~MeV; and uncertainties on photonic
corrections contribute 8~MeV.
The main improvements that are expected before the results are
finalised lie in the areas of the LEP beam energy, where a concerted 
programme is in progress to reduce the error, and the final-state 
interactions.

The basic physical problem which gives rise to the uncertainty over
final-state interactions is that when two W's in the same event both 
decay hadronically, the decay distance is smaller than typical
hadronisation scales. 
The hadronisation of the two systems may therefore not be 
independent, and so hadronisation models tuned to
\PZz$\to\mathrm{q}\overline{\mathrm{q}}$ decays may not properly
describe them.
Phenomenological models are used to study possible effects,
subdividing them into ``colour reconnection'' in the parton-shower
phase of the Monte Carlo models, and possible Bose-Einstein 
correlations between identical particles formed in the
hadronisation process.

\DOUBLEFIGURE{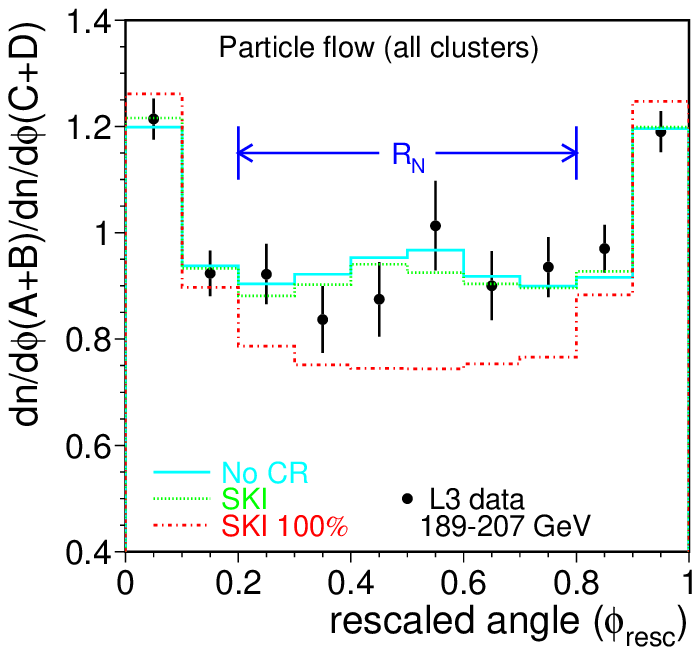}{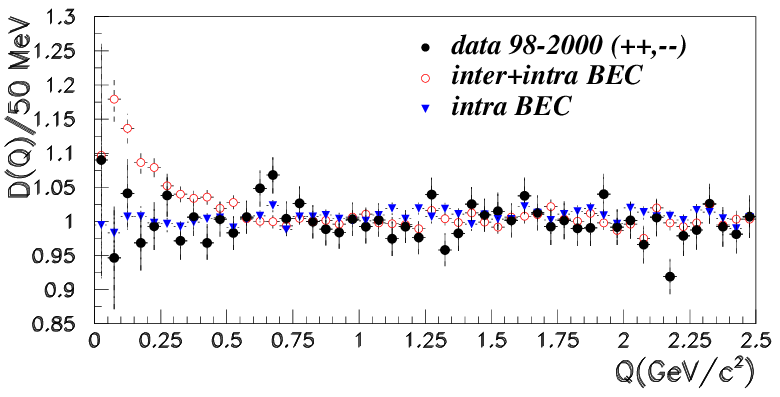}
{Particle flow ratio for particles between jets coming from the same
W relative to those between jets from different
W's\cite{bib:crl3}. Data are compared to models with no
colour reconnection, and to the SKI model\cite{bib:cr}.
\label{fig:cr}}
{Distribution of $D(Q)$ (see text) for DELPHI data, 
compared to different Monte Carlo
models\cite{bib:becdelphi}: ``intra'' refers to correlations within
one W decay, ``inter'' to those between W decays.
\label{fig:bec}}

A substantial effort has been spent in understanding
the possible effects of FSI models.
Recent work, in a collaborative effort between
all four LEP experiments, has focused on determining the common 
sensitivity to different models between different experiments, and on
developing ways to measure visible effects predicted by the models.

Sensitivity to the effect of colour reconnection models has been
obtained by
studying the particle flow between jets in $\Wqqqq$
events~\cite{bib:cr}. 
This is illustrated in Figure~\ref{fig:cr}. The data show some
sensitivity to the effects as predicted in the colour reconnection
models, and work continues to combine results from the four LEP
experiments to improve the sensitivity.

Bose-Einstein correlations are also being studied in
data~\cite{bib:bec}, in this case by comparing the two-particle
correlation functions, $\rho$, for
single hadronically decaying W's in $\Wqqlv$ events
($\rho^\mathrm{W}$),
and for $\Wqqqq$ events ($\rho^{\mathrm{W}\mathrm{W}}$). 
This may be expressed as~\cite{bib:chekanov}:
\[
\rho^{\mathrm{W}\mathrm{W}}(Q) = 2 \rho^{\mathrm{W}}(Q) + 
\rho_{mix}^{\mathrm{W}\mathrm{W}}(Q) + \Delta\rho(Q)
\]
where $\rho_{mix}^{\mathrm{W}\mathrm{W}}$ is evaluated from mixing
hadronic W decays from $\Wqqlv$ decays, and $\Delta\rho$ is any extra
part arising from correlations between particles from different W
decays in $\Wqqqq$ events.
Alternatively the ratio $D(Q)$ may be examined:
\[
D(Q) \equiv \frac{\rho^{\mathrm{W}\mathrm{W}}(Q)}{2 \rho^{\mathrm{W}}(Q) + 
\rho_{mix}^{\mathrm{W}\mathrm{W}}(Q)} .
\]
An observed $D(Q)$ distribution is shown in Figure~\ref{fig:bec}: a
deviation from unity at low $Q$ would most clearly signal the effect of
Bose-Einstein correlations between particles from different W's.
As illustrated in this figure, no evidence is observed of such an
effect.
As for colour reconnection, work is in progress to derive combined LEP
results in order better to constrain the possible effect on the W mass
measurement.

When the LEP measurement of $\mw$, given in Table~\ref{tab:lepmw} is
combined with that from \ppbar\ colliders as given in
Table~\ref{tab:mwhad}, a world average W mass of $80.451\pm0.033$~GeV
is obtained.
A similar combination of W width results gives
$\Gamma_{\mathrm{W}}=2.134\pm0.069$~GeV.

\section{Tests of the Gauge Couplings of Vector Bosons}

The gauge group of the Standard Model dictates the self-couplings of
the vector bosons, both in form and strength. 
The direct measurement of these couplings therefore provides a
fundamental test of the Standard Model gauge structure.
Electroweak gauge couplings have been measured directly at both LEP
and the Tevatron: at present constraints from LEP are 
more stringent.

W-pair production at LEP-2 involves the triple gauge
coupling vertex in two of the three lowest-order doubly-resonant
diagrams. 
Sensitivity to possible anomalous couplings is found in the W-pair
cross-section, and the W production and decay angle distributions.
Measurements have been reported at previous
conferences~\cite{bib:cctgcosaka}, but no combined LEP results have
been released recently because~\cite{bib:racoon,bib:kandy} 
higher-order corrections, previously neglected, are thought to be 
comparable to the current experimental precision~\cite{bib:villa}.

\EPSFIGURE{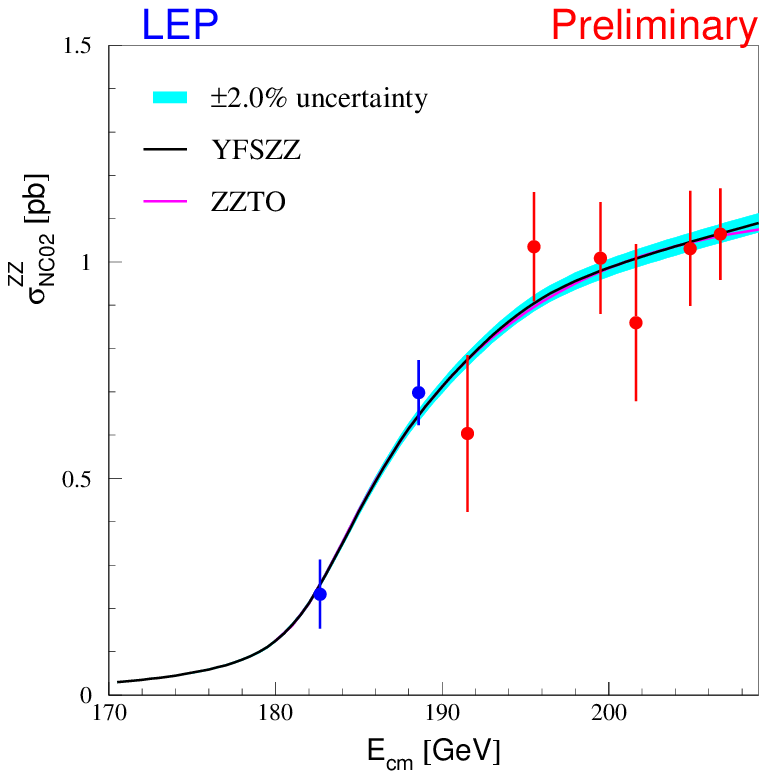}
{LEP averaged Z-pair production cross-section
measurements\cite{bib:sigzz} compared to Standard Model
predictions\cite{bib:smzz}.
\label{fig:sigzz}} 

Other measurements of triple gauge boson couplings are made at
LEP-2~\cite{bib:nctgc}
in the neutral vector boson 
processes of \PZz$\gamma$ and \PZz\PZz\ production. The
cross-section measured for the latter process is shown in
Figure~\ref{fig:sigzz} and is well-described by Standard Model
predictions. 
Measurements of quartic gauge couplings have also been made at LEP-2,
and were discussed in detail in other contributions to this 
conference~\cite{bib:qgchere}.

\section{Global Electroweak Tests}

Many of the individual results reported in preceding sections may be
used together to provide a global test of consistency with the
Standard Model.
If consistency with the model is observed, it is justifiable to go on
to deduce, in the framework of the
Standard Model, the unknown remaining parameter, the mass of the Higgs
boson, $\mh$.
The LEP electroweak working group has, for a number of years, 
carried out such global tests via a combined fit to a large number of
measurements sensitive to Standard Model parameters.
These results are reported here for the data available at this
conference.
These global fits use the electroweak libraries ZFITTER version
6.36~\cite{bib:zfitter} and TOPAZ0 version 4.4~\cite{bib:topaz0}
to provide the Standard Model predictions.
Theoretical uncertainties are included following detailed studies of
missing higher order electroweak corrections and their interplay with
QCD corrections~\cite{bib:precew}.
The precise LEP, SLD and Tevatron electroweak data are included, as
are $\sin^2\theta_W$ as measured in neutrino-nucleon (``$\nu$N'')
scattering\footnote{A new $\nu$N scattering result was
reported by the NuTeV Collaboration~\cite{bib:newnutev} during the
final stage of preparation of this contribution.
The $\sin^2\theta_{\mathrm{W}}$ result obtained differs from the
expected value by three standard deviations.}~\cite{bib:nuN}
and, new this year, atomic parity violation (``APV'') measurements in
caesium~\cite{bib:apv}.

Before making the full fit, the precise electroweak data from LEP and
SLD can be used together with $\alpha{(M_{\mathrm{Z}}^2)}$, the
$\nu$N and APV results to predict the masses of the top quark,
$\mtop$, and of the W, $\mw$.
The result obtained is shown in Figure~\ref{fig:mtopmw} by the solid
(red) contour.
Also shown are the direct measurements (dotted/green contour) 
of $\mtop=174.3\pm5.1$~GeV from the Tevatron~\cite{bib:mtop} 
and $\mw=80.451\pm0.033$~GeV obtained by combining LEP and \ppbar\
results; 
and the expected relationship between
$m_{\mathrm{W}}$ and $m_{\mathrm{top}}$ in the Standard Model for
different $\mh$ (shaded/yellow).
It can be seen that the precise input data predict values of $\mtop$
and $\mw$ consistent with those observed -- in both cases within two
standard deviations -- demonstrating that the
electroweak corrections can correctly predict the mass of heavy
particles.
For the W, the precision of the prediction via the Standard
Model fit is similar to that of the direct measurement.
For the top mass, the measurement is twice as precise as the
prediction.
It is observed in addition that both the precise input data and the
direct $\mw$/$\mtop$ measurements favour a light Higgs
boson rather than a heavy one.

\EPSFIGURE{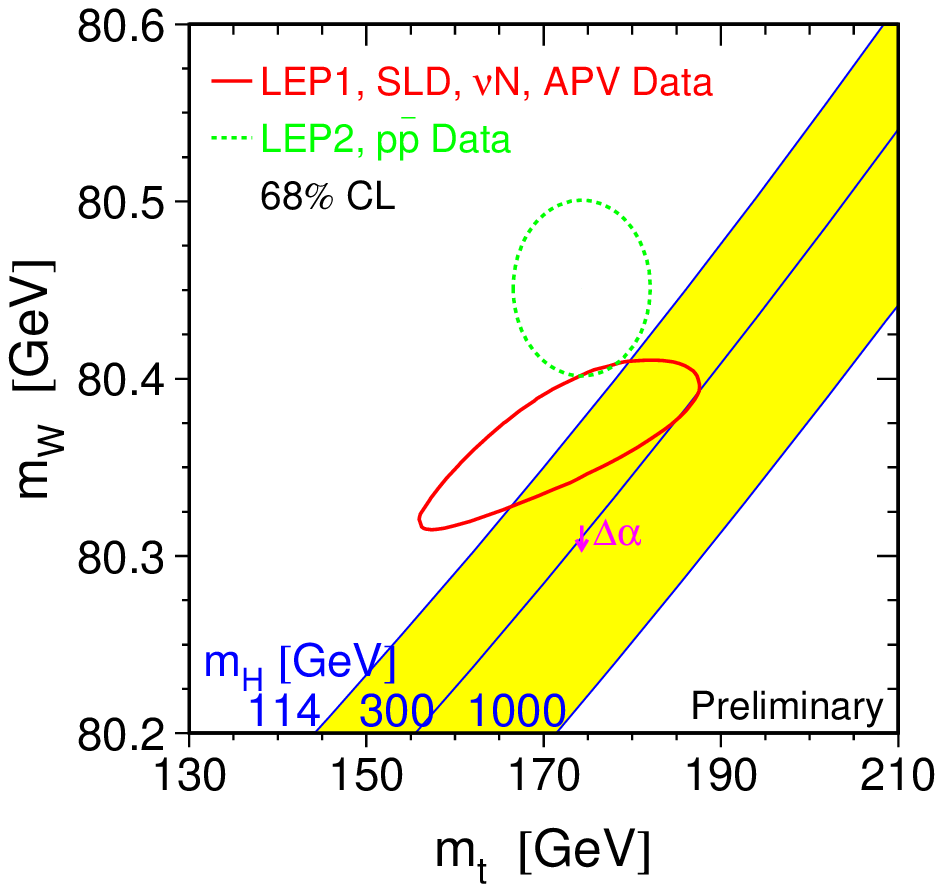}
{Comparison of direct and indirect constraints on the top and W
masses.
\label{fig:mtopmw}} 

\EPSFIGURE{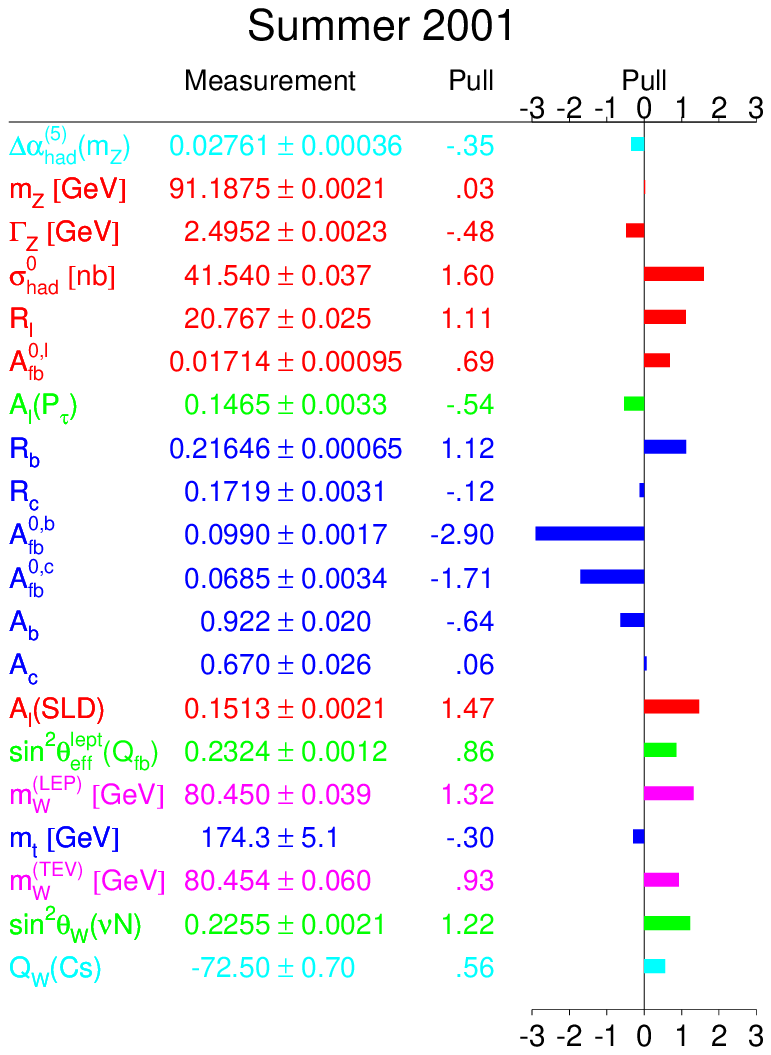}
{Pulls from the global electroweak fit.
\label{fig:pulls}} 

Going further, the full fit is made including also the $\mtop$ and
$\mw$ measurements.
The overall $\chi^2$ of the fit is 22.9 for 15 degrees of freedom,
corresponding to an 8.6\% probability.
To provide an impression of the contributions to this $\chi^2$, the 
best-fit value of each input datum is compared with the actual 
measurement, and the pull calculated as the difference between 
observation and best-fit divided by the measurement error. 
The results are shown in Figure~\ref{fig:pulls}.
The poorest description is of $\afbb$, which
is a reflection of the same disagreement discussed earlier in
Section~\ref{sec:asy}.
The best fit value of the Higgs mass is $\mh=88_{-35}^{+53}$~GeV,
where the error is asymmetric because the leading corrections depend
on $\log\mh$.
The variation above the minimum value of the $\chi^2$ as a function 
of the mass of the Higgs boson, $m_{\mathrm{H}}$, is shown in 
Figure~\ref{fig:blueband}.
The darker shaded/blue band enclosing the $\chi^2$ curve provides an
estimate of the theoretical uncertainty on the shape of the curve.
This band is a little broader than previously estimated because of the
inclusion of a new higher-order (fermionic two-loop) calculation of
$\mw$~\cite{bib:weiglein}. 
This has little effect via $\mw$ but does have an impact via
$\sintwl=\kappa_W(1-\mw^2/\mz^2)$.
This latter effect is controversial, and may well overestimate 
the true theoretical uncertainty, but it is currently included as 
equivalent two-loop calculations for Z widths and the effective
mixing angle are not available.
The $\chi^2$ curve may be used to derive a constraint on the Standard
Model Higgs boson mass, namely $m_{\mathrm{H}} < 196$~GeV at 95\% C.L.
Also shown in the Figure is the effect of using an alternative
theory-driven estimate of the hadronic corrections to
$\dahad(M_{\mathrm{Z}}^2)$~\cite{bib:martin} (dashed curve).
The effect on the $\mh$ prediction is sizable compared to the
theoretical uncertainty, for example.
The 95\% C.L. upper limit on $\mh$ moves to 222~GeV with this $\dahad$
estimate.

\EPSFIGURE{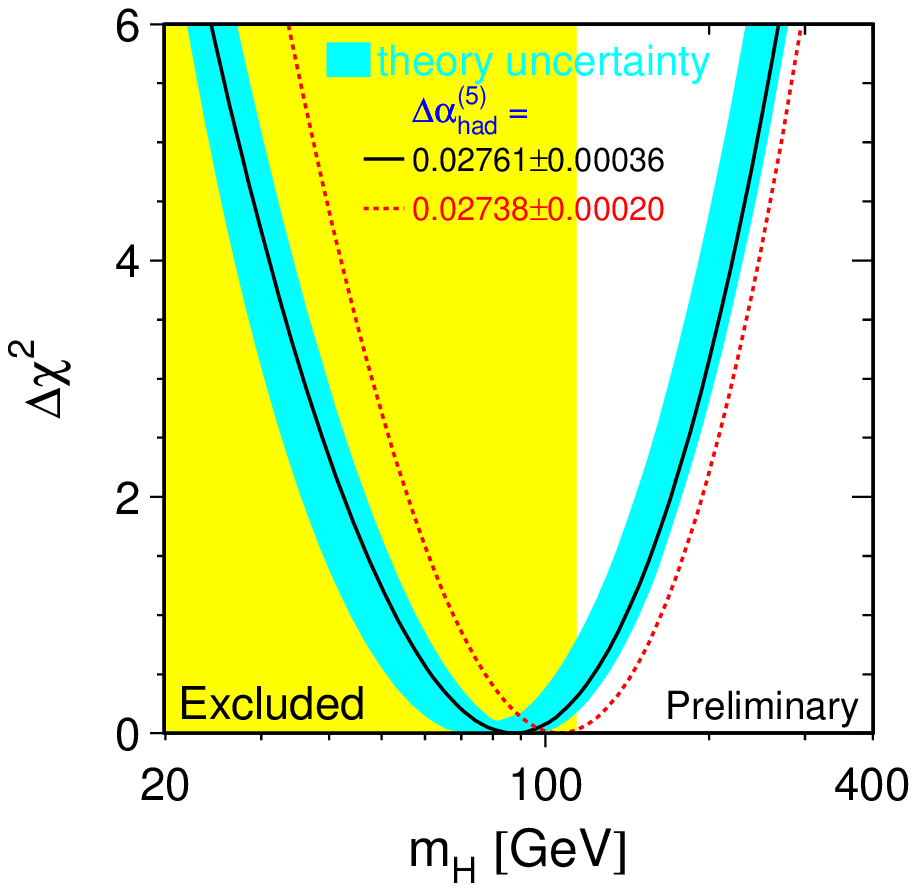}
{Constraining the Standard Model Higgs: $\Delta\chi^2$ curve as a
function of $\mH$ for the global electroweak fit.
The shaded band shows an estimate of the theoretical uncertainty, and
the lighter shaded area shows the region excluded by direct searches.
\label{fig:blueband}} 

\section{A Forward Look, and Conclusions}

The eleven years of data-taking by the LEP experiments, plus the
contributions of SLD, have
established that Standard Model radiative corrections describe
precision electroweak measurements. 
Data analysis is close to complete on the LEP-1 data, taken from
1989-1995. 
Work continues to finish LEP-2 analyses, and final results can be
expected over the next couple of years.
Improvements can still be expected in the W mass measurement, from
better understanding of final-state interaction effects in particular,
and in gauge-coupling measurements where the full data sample is not
yet included.

\EPSFIGURE{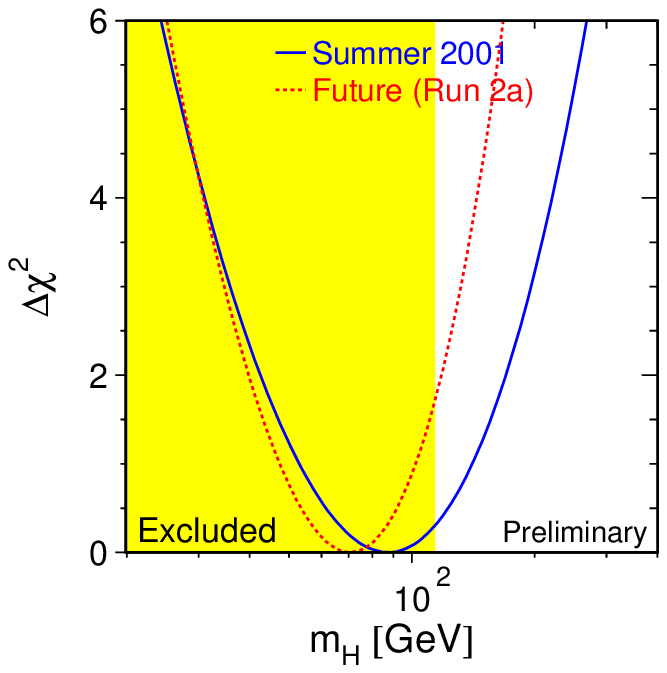}
{Typical $\Delta\chi^2$ curve obtained with $\mw$ and $\mtop$
measurements of 20~MeV and 3~GeV precision, respectively (dashed) and
compared to the current fit result central value (solid).
\label{fig:forward}} 

At the Tevatron, Run 2 data-taking has recently begun. 
Although luminosities are so far low, the expectation remains of
accumulating 2 fb$^{-1}$ in the next couple of years, which should allow
a W mass measurement with 30~MeV precision from each
experiment~\cite{bib:tevprospects}, and a top mass measured to
$\pm$3~GeV.
Combining the former result with the final $\mw$ results from LEP-2
should provide a world average W mass measurement error close to 20~MeV.
The effect such improvements could have, for example on the global fit
$\Delta\chi^2$ as a function of $\mh$, are shown in
Figure~\ref{fig:forward} (the central value of $\mh$ employed for the
future is, of course, arbitrarily selected).

Further substantial improvements in precision will have to wait for
the LHC and a future linear collider. The LHC should
improve the W and top mass precisions by a further factor two. The
main improvement would, of course, come from a discovery of the Higgs
boson, and a direct indication of whether it is the simplest
Standard Model particle.

In summary, precise tests of the electroweak sector of the Standard
Model have been made by a wide range of experiments, from the g-2
measurement in muon decays to LEP and the Tevatron.
Many of these tests have a high sensitivity to radiative corrections,
and the radiative correction structure is now rather well-established.
Two and three-loop calculations are essential in making
sufficiently precise predictions for some processes, and more progress
is still needed.
A small number of measurements, for example the measurement of
$\sintwl$ from the b forward-backward asymmetry at
LEP, show two or three standard deviation differences from
expectation which might point to possible cracks in the
Standard Model description, but none are compelling at present.
Further improvements in the quality of tests will arrive slowly
over the next few years: in particular further elucidation of the
electroweak symmetry-breaking mechanism will likely have to
await an improved discovery reach for a Higgs boson.

\section*{Acknowledgments}

The preparation of this talk was greatly eased by the work of the LEP
electroweak working group, and cross-LEP working groups on the W mass,
gauge coupling and fermion-pair measurements. 
In particular, I thank Martin Gr\"unewald for his unstinting help, and
Chris Hawkes for comments on this manuscript.
I also benefitted from the assistance of P. Antilogus, E. Barberio,
A. Bodek, D. Cavalli, G. Chiarelli, G. Cvetic, 
Y.S. Chung, M. Elsing, C. Gerber,
F. Gianotti, R. Hawkings, G.S. Hi, J. Holt, F. Jegerlehner, M. Kuze,
I. Logashenko, K. Long, W. Menges, K. M\"onig, A. Moutoussi,
C. Parkes, B. Pietrzyk, R. Tenchini, J. Timmermans, A. Valassi,
W. Venus, H. Voss, P. Wells, F. Yndurain and Z.G. Zhao.

\end{document}